\documentclass[sigconf]{acmart}
\usepackage[table]{xcolor}
\usepackage{xspace}
\usepackage{graphicx}
\usepackage{float}
\usepackage{multirow}
\usepackage{booktabs}
\usepackage{hyperref}
\usepackage[most]{tcolorbox}
\usepackage{listings}
\usepackage{xparse}
\usepackage{lipsum}
\usepackage{enumitem}

\NewDocumentCommand{\RQ}{m m o o}{%
    \textbf{RQ#1:} #2\\[0.5ex]
    \IfValueT{#3}{%
        {\hangindent=2em\hangafter=1
        \small{\textit{\textbf{Rationale:}} #3}\\[1.5ex]}
    }
    \IfValueT{#4}{%
        {\hangindent=2em\hangafter=1
        \small{\textit{\textbf{Evaluation:}} #4}\\[1.5ex]}
    }
    
}
\tcbset{
  colback=gray!10,    
  colframe=black!75,  
  boxrule=0.5pt,      
  arc=3pt,            
  left=4pt, right=4pt, top=4pt, bottom=4pt,
  fonttitle=\bfseries,
  enhanced,
  breakable
}

\newif\ifedit

\usepackage{xcolor}

\newcommand{\et}{\textit{et al.}\xspace}
\definecolor{purple1}{HTML}{8C4FFF}
\definecolor{purple2}{HTML}{7240D0}

\usepackage[skip=0pt,size=small]{caption}
\AtBeginDocument{%
  }

\copyrightyear{2026}
\acmYear{2026}
\setcopyright{cc}
\setcctype{by}
\acmConference[PROMISE '26]{21st International Conference on Predictive Models and Data Analytics in Software Engineering}{July 05, 2026}{Montreal, QC, Canada}
\acmBooktitle{21st International Conference on Predictive Models and Data Analytics in Software Engineering (PROMISE '26), July 05, 2026, Montreal, QC, Canada}
\acmDOI{10.1145/3803846.3807467}
\acmISBN{979-8-4007-2584-5/2026/07}

\acmBooktitle{Companion Proceedings of the 34th ACM Symposium on the Foundations of Software Engineering (FSE '26), June 5--9, 2026, Montreal, Canada}

\acmConference[PROMISE 2026]{The International Conference on Predictive Models and Data Analytics in Software Engineering (PROMISE)}{5 July, 2026}{Montreal, Canada}

\begin{document}
    
    \title{Beyond Code Snippets: Benchmarking LLMs on Repository-Level Question Answering}
    \author{Yoseph Berhanu Alebachew}
    \email{yoseph@vt.edu}
    \orcid{0000-0002-2922-4337}
    \affiliation{%
      \institution{Department of Computer Science, Virginia Tech}
      \city{Blacksburg}
      \state{Virginia}
      \country{USA}
    }

    \author{Hunter Leary}
    \email{hunterl22@vt.edu}
    \orcid{0009-0008-1056-8106}
    \affiliation{%
      \institution{Department of Computer Science, Virginia Tech}
      \city{Blacksburg}
      \state{Virginia}
      \country{USA}
    }
    \author{Swanand Vaishampayan}
    \email{swanandsv@vt.edu}
    \orcid{0009-0009-0664-8493}
    \affiliation{%
      \institution{Department of Computer Science, Virginia Tech}
      \city{Blacksburg}
      \state{Virginia}
      \country{USA}
    }
    
    \author{Chris Brown}
    \email{dcbrown@vt.edu}
    \orcid{0000-0002-6036-4733}
    \affiliation{%
      \institution{Department of Computer Science, Virginia Tech}
      \city{Blacksburg}
      \state{Virginia}
      \country{USA}
    }
    
    \renewcommand{\shortauthors}{Alebachew et al.}
    \thanks{AI was used to revise paragraphs in this manuscript.}

\begin{abstract}
Large Language Models (LLMs) have shown impressive capabilities across software engineering tasks, including question answering (QA). However, most studies and benchmarks focus on isolated functions or single-file snippets, overlooking the challenges of real-world program comprehension, which often spans multiple files and system-level dependencies. In this work, we introduce \textit{StackRepoQA}, the first multi-project, repository-level question answering dataset constructed from $1{,}318$ real developer questions and accepted answers across $134$ open-source Java projects. Using this dataset, we systematically evaluate two widely used LLMs (Claude 3.5 Sonnet and GPT-4o) under both direct prompting and agentic configurations. We compare baseline performance with retrieval-augmented generation methods that leverage file-level retrieval and graph-based representations of structural dependencies. Our results show that LLMs achieve moderate accuracy at baseline, with performance improving when structural signals are incorporated. Nonetheless, overall accuracy remains limited for repository-scale comprehension. The analysis reveals that high scores often result from verbatim reproduction of Stack Overflow answers rather than genuine reasoning. To our knowledge, this is the first empirical study to provide such evidence in repository-level QA.  We release StackRepoQA to encourage further research into benchmarks, evaluation protocols, and augmentation strategies that disentangle memorization from reasoning, advancing LLMs as reliable tool for repository-scale program comprehension.
\end{abstract}


\maketitle

\section{Introduction}
Program comprehension is a critical process for developing and maintaining software systems. However, modern codebases\footnote{In this paper we use the terms `project', `codebase', and `repository' interchangeably to refer to the entirety of the codebase as opposed to a single file or snippet.} are often too complex for developers to easily understand. As a result, comprehension is one of the most time-consuming aspects of software development. Studies estimate that professional developers spend about 58\% of their time trying to understand existing codebases~\cite{xia_measuring_2018}. 

In this process, developers often rely on asking questions as a key strategy ~\cite{xia_measuring_2018,latoza_importance_2010}, including seeking help on company-specific platforms or community-driven sites such as Stack Overflow\footnote{\url{https://stackoverflow.com}}~\cite{zhang2019reading}.

Compared to documentation---which is often outdated or poorly maintained---developers frequently perceive these social resources as more useful, timely, and reliable~\cite{roehm_how_2012,xia_measuring_2018}. However, maintaining question answering (QA) platforms is resource-intensive, and many questions remain unanswered or answers quickly become outdated as the underlying code evolves~\cite{zhang2019empirical}. This also slows open-source contributions, as potential contributors may not receive timely guidance when maintainers are overloaded\cite{alebachew2025are}.

Recent developments in Large Language Models (LLMs) show strong potential for a wide range of software engineering (SE) tasks, including program comprehension and QA~\cite{liu_large_2024, wei_emergent_2022}. By pretraining on vast internet-scale corpora, LLMs are rapidly becoming integral to developer workflows. Consequently, developers are increasingly using LLMs to answer questions they would have previously directed to peers or platforms like Stack Overflow~\cite{vaillant2024developers}. Surveys show that professionals rely on tools such as ChatGPT and Copilot to seek explanations, debug errors, and clarify concepts~\cite{liang2023large, zhang2023practices}.

Despite this potential, existing studies and datasets for LLM-based program comprehension---particularly in question-answering-style evaluations---predominantly focus on small-scale code snippets (e.g., single function or file)~\cite{xu_systematic_2022,zhang_unifying_2024}. This narrow scope limits the applicability of findings to real-world development, where the comprehension required to answer certain questions often spans multiple files and broader repository-level contexts. Prior efforts to move beyond single-file comprehension face limitations (see Section~\ref{sec:related_work}).

In the meantime, for repository‑level tasks, retrieval‑augmented generation (RAG) is designed to mitigate challenges of limited context size and the knowledge cutoff of LLMs. RAG pipelines address these challenges by selectively incorporating relevant external information into a model's limited context window. For QA tasks, RAG pipelines are expected to reflect the current state of the codebase rather than relying on project documentation, which---as noted earlier---is often incomplete or outdated~\cite{roehm_how_2012}.

In source-code RAG setups, retrieval solely based on flat textual similarity is insufficient, because code inherently encodes structural and semantic relationships---such as function calls, class hierarchies, and data-flow dependencies---that extend beyond what text similarity can capture~\cite{yin2024deep}. Moreover, indiscriminately providing the entire repository (when context window size permits) has been shown to be ineffective and can even degrade performance~\cite{rando_longcodebench_2025}, underscoring that repository-level QA requires deliberate exploration of code structure rather than brute-force context inclusion. 

To address this gap, we present \textbf{\textit{StackRepoQA}}---a dataset of $1{,}318$ real developer questions from Stack Overflow mapped to $134$ open-source Java repositories hosted on GitHub. Using this dataset, we investigate the following research questions.
\begin{enumerate}[label=\textbf{RQ\arabic*:},topsep=2pt]
    \item How accurately can LLMs answer repository-level developer questions, drawn from Stack Overflow, without any external augmentation?
    \item To what extent do RAG approaches---specifically file-level RAG and graph-based RAG---improve LLM performance on repository-level question answering?
\end{enumerate}

Our findings indicate that while LLMs achieve moderate accuracy (around $58\%$) on repository-level QA, much of this success can be attributed to memorization of previously seen Stack Overflow content rather than genuine reasoning over source code. RAG provided measurable improvements, with graph-based retrieval yielding the largest gains; however, even the best configuration only increased accuracy to approximately $64\%$.

Performance was consistently lower on questions posted after the models' training cutoff dates, confirming that strong results on publicly available repositories do not necessarily reflect an ability to reason about unseen or evolving projects. Collectively, these findings underscore both the limitations of current LLMs for repository-level comprehension and the potential of structured, graph-based retrieval as a step toward more robust and reliable solutions.

\subsubsection*{\textbf{Contributions}}  
\noindent
To our knowledge, this is the first work to introduce a multi-project, repository-level question answering dataset derived from real developer questions with accepted answers, and the first to incorporate code structure into LLM-based QA. Our study highlights the limitations of current LLMs in realistic QA scenarios and shows that structured retrieval can partially mitigate these challenges. Specifically, this data-oriented paper makes the following key contributions:
\begin{itemize}[left=1em,topsep=2pt]
    \item \textbf{StackRepoQA:} We release the first publicly available, multi-project dataset for repository-level Q\&A-pairs, containing $1{,}318$ \textbf{real developer} questions mapped to $134$ actively maintained Java projects. This curated dataset and evaluation pipeline are made available for the community to use.
    
    \item \textbf{Systematic Evaluation:} We evaluate two widely used LLM models on repository-scale QA and introduce a multi-agent RAG framework that integrates file-level retrieval and graph-based structural queries for controlled comparison of augmentation strategies. 
    
    \item \textbf{Ablation Study:} We show that graph-level retrieval provides the largest performance gains and analyze memorization effects using pre- and post-cutoff questions to quantify the influence of training data exposure versus genuine reasoning.
    
    \item \textbf{Insights and Resources:} We discuss key limitations of LLMs on repository-level tasks, such as performance degradation on unseen questions and sensitivity to irrelevant context.
    
\end{itemize}

\noindent \subsubsection*{\textbf{Implications}.}

Our findings highlight the need for cutoff-aware benchmarks and structure-aware retrieval, while cautioning practitioners against relying on LLMs as stand-alone tools for repository-level comprehension in evolving codebases.

\section{Related Work}\label{sec:related_work}
\subsection*{\large Repository-Level Question Answering}
Repository-level activities like program comprehension differ from their snippet-level counterparts in that they require developers to build mental models across files, modules, and heterogeneous artifacts. Prior studies show that developers rely on structural cues for system-level reasoning~\cite{roehm_how_2012} and adapt strategies across conceptual, debugging, and integration tasks~\cite{latoza_importance_2010}.

Recent work on LLMs for SE moves beyond snippet-level tasks toward repository-grounded question answering (QA). \emph{SpyderCodeQA}~\cite{strich_improving_2024} provides manually curated question-answer pairs (Q\&A-pairs) for the Spyder IDE that probe code semantics, dependencies, and meta-information, but its single-project scope limits generalizability. Moreover, its approach relies on fine-tuning a model for each project, reducing practicality. 

\emph{CoReQA}\footnote{Dataset not publicly available}~\cite{chen_coreqa_2025} and \emph{CodeRepoQA}~\cite{hu_coderepoqa_2024} present issue-derived datasets where an LLM reconstructed questions and answers from GitHub issues and discussions across multiple repositories. However, because the Q\&A pairs are LLM generated rather than directly obtained from developers, these datasets do not reflect questions posed by real developers along with their accepted answers, limiting both realism and reproducibility.

In contrast, our work introduces the first publicly available, multi-project dataset for repository-level question answer pairs derived from real developer-authored Stack Overflow questions with accepted answers. By explicitly supporting temporal cutoff analysis, StackRepoQA enables disentangling memorization from reasoning—an aspect not addressed by prior resources. 

\subsection*{\large Long Context vs. Retrieval Augmentation}
Long-context models promise end-to-end reasoning over large repositories, but multiple studies show that performance often degrades as context length increases, and even when the relevant code is present, models often fail to locate it reliably~\cite{rando_longcodebench_2025}. This is partly because longer input reduces an LLM's ability to capture dependencies across the context~\cite{rando_longcodebench_2025}. Moreover, source code relationships are not strictly linear. Cross-file dependencies and structural connections must be explicitly reasoned about.

RAG has therefore become a common strategy for grounding LLM responses~\cite{lewis2020retrieval,izacard2022atlas}. However, recent work shows that irrelevant or adversarially ``hard'' negatives can significantly reduce accuracy, even when relevant passages are included~\cite{heydari2024context}. This phenomenon, sometimes called the \emph{distracting-context effect}, highlights the vulnerability of LLMs to noisy or misleading retrieval and motivates adaptive filtering or gating strategies.

To overcome the limitations of long-context beyond flat text-based retrieval, structure-aware pipelines have been proposed that index repositories as graphs (e.g., code property graphs, symbol graphs). Such approaches enable multi-hop queries across cross-file relations and consistently outperform flat retrieval on code tasks~\cite{liu2024codexgraph,phan_repohyper_2024,cheng_dataflow-guided_2024}. 




In summary, prior research establishes two key themes: (i) repository level QA is qualitatively harder than snippet-level tasks; and (ii) longer context windows alone do not solve retrieval and grounding. Our work advances the state of the art by (a) releasing a \emph{public}, multi-project, cutoff-aware benchmark derived from real developer questions with accepted answers, and (b) empirically showing that structure-aware (graph) retrieval provides consistent gains over file-only RAG while exposing limitations of current LLMs for project-scale comprehension.

\section{Methodology} 
\subsection{Dataset Construction}
\begin{figure}[ht]
    \centering
    \includegraphics[width=0.85\linewidth]{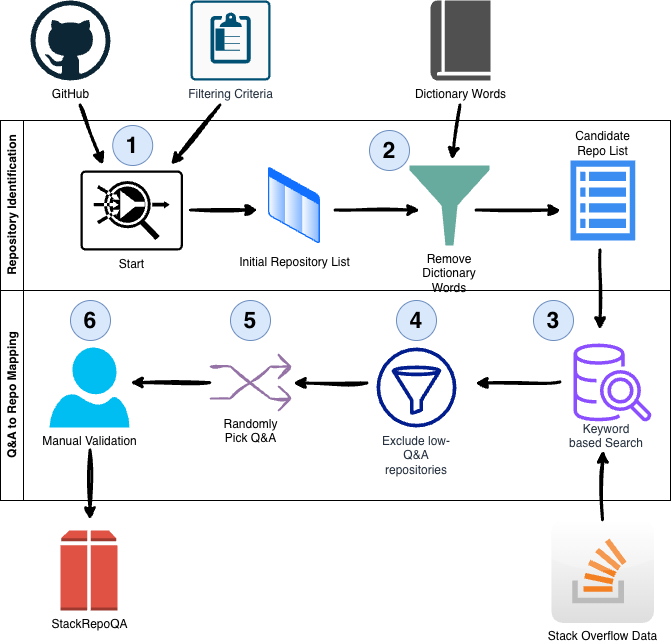}
    \caption{Overview of the data collection and preprocessing pipeline.}
    \Description{Block diagram illustrating the StackRepoQA data collection pipeline, including GitHub repository selection, Stack Overflow question and answer extraction, filtering, and mapping questions to source code artifacts.}
    \label{fig:data_collection}
\end{figure}

This section describes the construction of the StackRepoQA dataset. Figure~\ref{fig:data_collection} provides an overview of the complete data collection and processing pipeline, including repository selection, Q\&A-pairs extraction, and the mapping of developer questions to source code artifacts. Each stage is detailed in the following subsections.

\subsubsection*{\textbf{GitHub Repositories:}}  
We used the GitHub API\footnote{\url{https://docs.github.com/en/rest/about-the-rest-api/about-the-rest-api}} to collect metadata for public Java repositories that are both \textbf{large} and \textbf{active}. Data were collected on June~9,~2025 at 8:45~PM~EST. Repositories were ranked by GitHub star count, a widely used proxy for popularity and community interest~\cite{borges2016understanding}. Because GitHub hosts a large number of personal, toy, or inactive repositories, careful filtering is required to identify engineered, production-level projects~\cite{kalliamvakou2014promises}. We therefore adapted the criteria of Kalliamvakou et al.~\cite{kalliamvakou2014promises}, requiring that each repository have at least 20 contributors, at least 10 GitHub stars, at least 1,000 commits, a minimum size of 10MB, a permissive license (e.g., MIT, New BSD, Apache-2.0), and at least one commit within the 12 months prior to data collection.


We restricted our search to repositories where Java is the primary language, motivated by its widespread use in large-scale industrial and open-source systems~\cite{lu2020analysis}. While our dataset focuses on Java, the methodology is language-agnostic and can be applied to other ecosystems.

We further manually excluded repositories that did not contain source code (e.g., documentation-only projects, tutorials, or curated resource lists). To reduce ambiguity when mapping repositories to Stack Overflow questions, we also removed projects whose names coincide with common dictionary words (Step~2 in Figure~\ref{fig:data_collection}). This process yielded a final set of $420$ actively maintained Java repositories, representative of realistic, production-level software systems.


\subsubsection*{\textbf{Stack Overflow Questions:}}  
We obtained the archived version of Stack Overflow questions and answers from the official data dump~\cite{stackoverflow_dump2025}. The archive contains all content posted on \url{stackoverflow.com} up to April~3,~2025. We chose Stack Overflow due to its role as a primary question-answering platform for software developers, where questions are authored by practitioners encountering concrete problems and answers are vetted through community feedback and acceptance mechanisms~\cite{treude2011programmers,barua2014developers}. Prior studies have shown that Stack Overflow questions capture realistic information needs during program comprehension and maintenance, making them a reliable proxy for real-world developer questions~\cite{treude2011programmers}.

In contrast to recent datasets (e.g., CoReQA, CodeRepoQA) that reconstruct or rephrase question and answer pairs (Q\&A pairs) from GitHub issues using LLMs, our approach preserves the original, developer-authored questions and accepted answers. LLM-generated Q\&A pairs risk introducing artifacts and implicit solution knowledge, obscuring how developers naturally formulate problems and confounding memorization with reasoning. By grounding StackRepoQA in organically produced Stack Overflow Q\&A pairs with accepted answers, we prioritize realism, reproducibility, and faithful representation of developer information needs.

\subsubsection*{\textbf{Q\&A pairs To Repository Mapping:}}  
To construct StackRepoQA, we mapped GitHub repositories to Stack Overflow questions by performing a keyword-based search using each repository's name (Step~3 in Figure~\ref{fig:data_collection}). This process yielded $233{,}175$ candidate questions. From this pool, we randomly sampled seven to ten questions per repository to satisfy the statistical power requirements described in Section~\ref{sec:experiment}. We restricted sampling to questions with accepted answers, ensuring that each question is paired with a response that was endorsed by the Stack Overflow community.

We subsequently excluded repositories (Step~4) that did not meet a minimum threshold of seven associated question--answer pairs. This threshold was informed by a power analysis~\cite{cohen_statistical_1988}, which indicated that at least seven observations per project are required to support paired statistical analyses and repository-level comparisons. To mitigate potential false positives arising from the keyword-based mapping strategy, we additionally retrieved up to three extra candidate questions per repository when available. Following this process, we obtained a total of $1{,}318$ question--answer pairs mapped to $134$ GitHub repositories. Table~\ref{tab:repo_statistics} presents summary statistics for the final list repositories retained after filtering. 
\begin{table}[htbp!]
    \centering
    \footnotesize
    \caption{Basic statistics on dataset repositories after filtering}
    \resizebox{0.47\textwidth}{!}{
    \rowcolors{1}{purple1!10}{white}
            
        \begin{tabular}{lrrrrr}
            \toprule
             \rowcolor{purple1!60} & \color{white} \textbf{No. Forks}  & \color{white} \textbf{Size (MB)} & \color{white} \textbf{Stars} & \color{white} \textbf{Contributors} & \color{white} \textbf{Age In Years} \\
            \midrule
            \textbf{Average} & 1954.51 & 264.79 & 7220.50 & 176.58 & 11.62 \\
            \textbf{Max} & 13581.00 &  5777.54 & 57793.00 & 411.00 & 16.65 \\
            \textbf{Median} & 984.00 & 110.37 & 3741.00 & 148.00 & 11.78 \\
            \textbf{Min} & 63.00 & 10.71 & 958.00 & 21.00 & 2.218 \\
            \textbf{std($\sigma$)} & 2350.26 & 562.66 & 8751.16 & 110.08 & 2.96 \\
            \textbf{Total} & 259950 & 35217.55 & 960327 & 23485 & 1545.21 \\
            \bottomrule
        \end{tabular}
    }
    \label{tab:repo_statistics}
\end{table}
To identify and label false positives, a software developer with over 15 years of professional Java experience manually reviewed each question to assess whether it was genuinely related to the mapped repository. This inspection identified $n = 249$ questions that were deemed unrelated (e.g., cases where the repository name appeared in an unrelated context). Rather than removing these items, we retain them in the dataset with explicit markings, enabling their use as control cases in our analyses and in future work. To assess annotation reliability, two additional raters independently reviewed a random subset of 20\% of the questions, resulting in a perfect agreement which indicates high annotation consistency.

Finally, we note two important characteristics of the resulting dataset. First, while our data collection process does not independently verify the technical correctness of answers, we adopt the presence of an accepted answer as a practical proxy for correctness, a convention widely used in prior empirical studies leveraging Stack Overflow data~\cite{zhang2019empirical, kavaler2018determinants}. Although accepted answers are not guaranteed to be universally correct or optimal, they reflect community validation and the satisfaction of the original question asker. Second, we do not claim that all questions in StackRepoQA require repository-level understanding to answer. Instead, the dataset intentionally reflects the heterogeneous nature of real-world developer questions, ranging from localized API usage and configuration issues to questions that demand reasoning across multiple files, components, or external artifacts. This diversity mirrors authentic developer workflows and enables evaluation across varying degrees of contextual and structural complexity. For instance, the following question from our dataset: ``\textit{...I'm trying to insert data into Cassandra table using this code in Java in REST server created by DropWizard...\texttt{<code>}...But it gives me this error:...\texttt{<error message>} What do you think is the problem?}''\footnote{\url{https://stackoverflow.com/questions/44910097/cassandra-insertion-invalidqueryexception-invalid-null-value-in-condition-for}} requires knowledge of the Java source code, Apache Cassandra configurations, the DropWizard framework, and their interactions within the project.
    

\subsection{Experiment}\label{sec:experiment} 
\subsubsection*{\textbf{Sampling:}} 
To assess the ability of LLMs to answer repository-level questions, we conducted an empirical evaluation using StackRepoQA. Although the dataset contains $1{,}318$ question–answer pairs, resource constraints prevented evaluation on the full set. We therefore performed a power analysis~\cite{cohen_statistical_1988,faul_gpower_2009} to determine the minimum number of samples required to detect statistically significant effects, including per-project differences. Such sampling based on power analysis is common practice in empirical SE research to ensure adequate statistical validity under resource constraints~\cite{arcuri_practical_2011}. Based on this analysis, we selected $482$ questions spanning $49$ repositories.

We conducted paired, within-question comparisons across conditions and analyzed scores using non-parametric paired tests alongside a linear mixed-effects model (condition as a fixed effect; random intercepts for question and project). Because score distributions violated normality assumptions, we relied on rank-based tests and reported \textit{Cliff's $\delta$} as the primary effect size measure. Cliff's $\delta$ is a distribution-free effect size commonly used with non-parametric tests such as the Wilcoxon signed-rank test, and it quantifies the probability that a randomly selected score from one condition exceeds a randomly selected score from another condition. This makes it particularly suitable for our ordinal, bounded (1--10) evaluation scores and paired experimental design. The observed between-condition effect magnitude (Cliff’s $\delta = 0.396$, corresponding approximately to Cohen's $d \approx 0.74$) falls in the moderate range and is readily detectable given our paired sample size.

\subsubsection*{\textbf{Choice of LLMs:}}
We evaluated two widely used LLMs for the task of repository-level question answering: \textbf{Claude 3.5 Sonnet} (June 2024 release) by Anthropic and \textbf{GPT-4o} (May 2024 release) by OpenAI. Recent industry surveys indicate that OpenAI's GPT models and Anthropic's Claude Sonnet are among the most widely used LLMs by software developers. According to Stack Overflow's 2025 Developer Survey, approximately 81–82\% of respondents reported using GPT models, while around 43\% reported using Claude Sonnet~\cite{stackoverflow2025survey_usage}. These findings suggest that GPT and Claude models are widely adopted in practice and suitable baselines for code-oriented LLM evaluations in empirical studies.  

In addition, GPT-4o exhibited the highest performance on the CoReQA benchmark~\cite{chen_coreqa_2025}. Claude 3.5 Sonnet has also demonstrated strong capabilities in structured reasoning; for example, it achieved the highest execution accuracy on CypherBench, a benchmark for natural language–to–Cypher translation on Neo4j\footnote{\url{https://neo4j.com/}} graph databases~\cite{yanlin2024cypherbench}. This was particularly important for the Graph RAG Agent (described later) which utilized Neo4j. For both LLMs, we used default inference settings to reflect typical developer usage. 
\subsubsection*{\textbf{Experimental Setup:}}
We first prompted the LLMs to answer Stack Overflow questions directly, without access to any external tools or augmentation. This \textit{baseline} reflects the models' inherent knowledge, acknowledging that both Stack Overflow content and public GitHub repositories are likely to be included in their training data.

Beyond this baseline, we implemented an \textbf{agentic setup} in which a supervisor agent interacts with one summarizer and two specialized retrieval augmented agents: a file-based RAG agent that retrieves relevant code fragments from a local copy of the repository, and a graph-based RAG agent that leverages structural dependencies stored in a Neo4j database. This framework allows the models to ground their answers in repository-specific context rather than relying solely on pretrained knowledge. 

Recent work in SE highlights the benefits of agentic systems. For example, Mazur \et~\cite{mazur2025querying} show that LLMs equipped with tool access achieve accuracy comparable to direct prompting while substantially reducing token usage. Similarly, Robeyns \et~\cite{robeyns2025self} demonstrate a self-improving coding agent that iteratively refines its behavior using external tools, resulting in 17–53\% performance improvements on SWE-Bench Verified benchmarks---far exceeding static LLM baselines. These findings support our design choice: agentic systems, with planning, tool integration, and feedback loops, are more effective than standalone LLMs for complex SE tasks.

\begin{figure}[ht]
    \centering
    \includegraphics[width=0.98\linewidth, height=5cm]{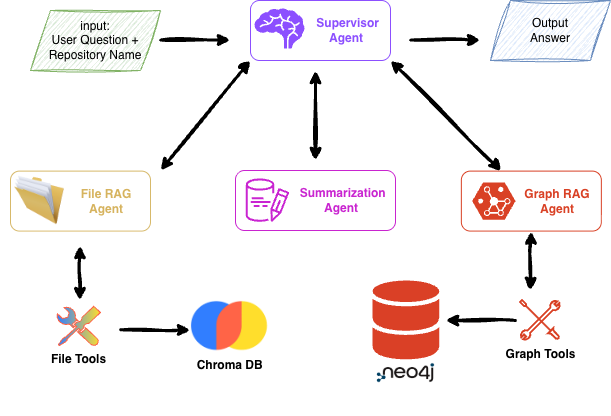}
    \caption{Overview of the multi-agent architecture used in our evaluation}
    \Description{Block diagram showing the multi-agent architecture employed in the experimental setup, including file-level and graph-level retrieval agents, summarization agent, and supervisor agent.}
    \label{fig:agent_architecture}
\end{figure}

\subsubsection*{\textbf{Architecture}:}
Our agentic setup consists of two specialized retrieval agents and two utility agents integrated with the LLM. For the experiment, the vector and graph database were pre-populated with repository data:
\begin{itemize}[left=1em]
    \item \textbf{File-RAG Agent:} Performs dense retrieval over a vector store (i.e., ChromaDB) of source code files, embedded using the \texttt{all-mpnet-base-v2} model~\cite{reimers2019sentencebert}. Given a natural language question, it retrieves the top-$k$ most semantically relevant files or code fragments using cosine similarity. We set $k=5$ to balance precision and recall, consistent with prior RAG studies~\cite{lewis2020retrieval, strich2024on}. This agent grounds the LLM in file-level context, which has been shown to improve answer quality in SE tasks.

    \item \textbf{Graph-RAG Agent:} Queries a Neo4j graph database constructed from the repository's abstract syntax tree (AST) and static dependency relations. The graph encodes entities such as classes, methods, and fields, along with relationships (\texttt{CALLS}, \texttt{CONTAINS}, \texttt{EXTENDS},  \texttt{USES}). For a given user query, the agent generates Cypher queries to retrieve subgraphs that capture relevant structural and cross-file dependencies. This approach is motivated by prior work showing that LLMs augmented with graph-based retrieval improve performance on repository-level tasks~\cite{liu2024codexgraph}.
    
    \item \textbf{Summarization Agent:} Synthesizes a final response from the outputs retrieved by the information agents. It produces a coherent, context-rich answer tailored to the user's question. In setups where the LLM's internal knowledge is used, this agent integrates it with the retrieved context to enhance completeness and fluency.

    \item \textbf{Supervisor Agent:} Coordinates the execution flow across all agents. It manages query routing, determines the order of information agent invocations, and passes their outputs to the summarization agent. The supervisor ensures consistency across the multi-agent pipeline and allows modular experimentation with different agent configurations. Each retrieval agent could be invoked to a maximum of three rounds by the supervisor. 
\end{itemize}

We implemented this architecture, illustrated in Figure~\ref{fig:agent_architecture}, using the \texttt{AutoGen} framework\footnote{\url{https://github.com/microsoft/autogen}} (version 0.9.6), and include the complete source code and outputs from all experimental runs in the replication package\footnote{anonymous2026supplementary}.

\subsubsection*{\textbf{Prompt Design:}} 
We structured the system prompt around five key components—Role, Context, Task, Rules, and Output—following the ``prompt canvas'' framework of Hewing and Leinhos~\cite{hewing_prompt_2024} and the component-based template analysis of Mao \et~\cite{mao_prompts_2025}. The specific prompts used for each agent are provided in the replication package~\cite{anonymous2026supplementary}.

Questions are first presented to the LLM. In the baseline evaluation, the LLM generates responses without any augmentation. In the agentic setup, the LLM routes the query through one or both retrieval agents. The retrieved context---either relevant files or graph substructures---is appended to the prompt and returned to the LLM to produce a grounded answer. We evaluate three configurations: (1) \textit{LLM only}; (2) \textit{LLM + File-RAG}; and (3) \textit{LLM + File-RAG + Graph-RAG}.

\subsubsection*{\textbf{Ablation Study:}}
To evaluate the contribution of each information agent—namely, File-RAG and Graph-RAG—we conducted an ablation study by individually disabling each agent. We also assessed two configurations in which all agents were enabled. 

\subsection{Evaluation and Analysis}
\subsubsection*{\textbf{LLM-as-judges:}}
Given that our primary goal is to evaluate the semantic and factual similarity between the model-generated responses and reference answers, we use LLM-as-judges~\cite{wang_can_2025} instead of traditional metrics like BLEU~\cite{papineni2002bleu}, ROUGE~\cite{lin2004rouge}, or edit similarity~\cite{navarro2001guided}. Traditional metrics mainly focus on surface-level overlap such as exact word matches or phrase similarity, which do not reliably capture semantic equivalence, especially when paraphrasing or varied wording occurs \cite{novikova2017need}. Additionally, using LLMs as judges allows for a more nuanced and human-like evaluation of meaning, accounting for context and subtle variations in expression \cite{wang_can_2025}. This ensures that our evaluation aligns more closely with our goal of measuring meaningful similarity rather than mere linguistic similarity.

In the LLM-as-a-judge setup, we presented the LLM with questions from StackRepoQA along with reference and candidate answers, asking it to rate responses (on correctness, completeness, faithfulness, conciseness) between 1–10---mirroring methodologies from prior studies that employed LLMs as judges in QA and software engineering evaluations~\cite{liu2023geval,lin2022truthfulqa,tong2024codejudge,wang_can_2025, bose2025llm}.

The evaluation prompt provided to the LLM-as-a-Judge (see replication package~\cite{anonymous2026supplementary}) included a fixed scaffold with a clear rubric and scoring guide. Aside from the ablation study, each response was evaluated independently by two LLMs to mitigate the self preference bias observed in prior work, where models tend to assign higher scores to their own outputs compared to those of competing models~\cite{zheng2020judging}.

Two experienced software engineers evaluated 30 random responses and we calculated inter-rater agreement between two human judges and the LLM. Pairwise weighted Cohen's $\kappa$~\cite{cohen1960coefficient} indicated substantial agreement between Human~1 and Human~2 ($\kappa=0.78$), and comparable levels of agreement between the LLM and Human~1 ($\kappa=0.77$) as well as almost perfect agreement between the LLM and Human~2 ($\kappa=0.87$). 

Following the interpretation guidelines of Landis and Koch~\cite{landis1977measurement}, these values indicate substantial agreement.  When considering all three raters jointly, the Intraclass Correlation Coefficient (ICC)~\cite{shrout1979intraclass,koo2016guideline} further confirmed high reliability. Specifically, $ICC(2,1)=0.82$, 95\% CI [0.69, 0.90], $p<.001$, indicated substantial reliability for a single rater, while $ICC(2,k)=0.93$, 95\% CI [0.87, 0.96], $p<.001$, indicated excellent reliability when averaging across raters. These results demonstrate that the LLM's evaluations align closely with human judgments and can serve as a reliable complementary evaluator.

\subsubsection*{\textbf{Analysis}}
To address our first research question, we compared the mean scores of responses across the different configurations and LLMs. Prior to hypothesis testing, we assessed the distribution of the scores using the Shapiro--Wilk test~\cite{shapiro1965analysis}. Since the results indicated non-normality, we employed the non-parametric Wilcoxon Signed-Rank test to assess statistical significance of pairwise differences. For effect size estimation under non-normal conditions, we calculated Cliff's delta ($D$), a robust, distribution-free measure of effect size~\cite{cliff1993dominance}.

In addition, we conducted a \textit{cutoff analysis} to investigate whether the posting date of the Stack Overflow question affected model accuracy. Foundational LLMs are trained on corpora up to a fixed cutoff date during pretraining~\cite{carlini2021extracting}. Questions and their corresponding answers posted before this date are likely to have been included in the training data, while those posted afterwards should be unseen by the model. This analysis allowed us to disentangle performance due to memorization from performance due to genuine reasoning. Finally, we used Spearman correlation coefficients to analyze trends based on repository and project characteristics.

\section{Results}\label{sec:results}

\subsection{RQ1: LLM Accuracy}
The first research question addresses how well LLMs can answer Stack Overflow questions without any form of augmentation. As depicted in Table \ref{tab:llm_results}, our results show that, on average, GPT-4o (\textbf{\textit{avg. score = 6.94}}) outperformed Claude 3.5 Sonnet (\textbf{\textit{avg. score = 6.67}}) for all questions (i.e., related + unrelated). This trend holds both when each LLM evaluated its own responses (GPT: \textbf{\textit{7.08}}, Claude: \textbf{\textit{5.88}}) and when models cross-evaluated each other's outputs (GPT scoring Claude: \textbf{\textit{7.21}}; Claude scoring GPT: \textbf{\textit{6.77}}). 

\begin{table}[ht]
    \centering
    \caption{Average scores from different LLM run without any augmentation and both LLMs as judges}
    \label{tab:llm_results}
    \rowcolors{1}{purple1!10}{white}
    \resizebox{0.47\textwidth}{!}{
        \begin{tabular}{|c|c|c|c|c|c|c|c|}
            \hline
            \rowcolor{purple1!60} \color{white} 
        \multirow{2}{*}{} & 
            \multicolumn{3}{c|}{ \color{white} \textbf{Related Questions}} & \multicolumn{3}{c|}{ \color{white} \textbf{Unrelated Questions}} &  \color{white} \textbf{All}\\
            \cline{2-8}
            \rowcolor{purple1!60} 
             \color{white} \textbf{LLM} &  \color{white} \textbf{Claude} &  \color{white} \textbf{GPT} &  \color{white} \textbf{Average} &  \color{white} \textbf{Claude} & \color{white} \textbf{GPT} &  \color{white} \textbf{Average} &  \color{white} \textbf{Average} \\
            \hline
            Claude & 5.88  & 7.21  & 6.54 & 6.97 & 7.99 & 7.48 & 6.67\\ 
            \hline
            GPT-4o & 6.77 & 7.08 & 6.93 & 6.96 & 7.13 & 7.06 & 6.94\\
            \hline
            \multicolumn{3}{|c|}{\textbf{Average}} & \textit{\textbf{6.73}} & \multicolumn{2}{c|}{} & \textit{\textbf{7.27}} & \textit{\textbf{6.81}} \\
            \hline
        \end{tabular}
    }
\end{table}

Interestingly, responses to questions unrelated to the mapped project slightly outperformed those related to the project (i.e., \textbf{\textit{7.27}} vs.\ \textbf{\textit{6.73}}). While this difference was not statistically significant, it suggests that the models performed consistently across both categories. However, this gap widened considerably when we prompted the LLMs to ignore their inbuilt knowledge and instead rely solely on information retrieved through external tools and agents, following established open-book and faithful question answering protocols~\cite{lewis2020retrieval,dziri2022faithdial}. This pattern indicates that much of the higher performance without augmentation likely stems from the models' prior exposure to the question--answer pairs during training, rather than from genuine reasoning over large codebases.


As part of the cutoff analysis, we partitioned questions into those asked before and after the LLM's cutoff date (April 20, 2024 for Claude 3.5 Sonnet~\cite{anthropic2025cutoff} and June 1, 2024 for GPT-4o~\cite{openai2025cutoff}). We observed a significant drop in performance after the cutoff for both models: Claude's average score decreased from 6.59 to 5.44 (p = 0.0009), while GPT-4o's dropped from 6.97 to 5.97 (p = 0.0024), highlighting the influence of memorized training data on their repository-level question-answering capabilities.



Finally, we examined whether repository-level characteristics such as size, activity, or popularity correlated with model performance. Spearman correlation coefficients were uniformly small (e.g., correlations with project size ranged from $-0.24$ to $-0.07$, and popularity metrics such as stars and forks remained below $|0.22|$), indicating that these characteristics do not systematically influence LLM accuracy in repository-level question answering.

\begin{tcolorbox}[colback=purple1!10, colframe=purple2!60, title=Key Finding (RQ1)]
Without external augmentation, LLMs achieve only moderate accuracy on repository-level question answering. Performance degrades sharply on post-cutoff questions, indicating that much of the apparent success on earlier questions is driven by memorization rather than repository-level reasoning.
\end{tcolorbox}


\subsection{RQ2: Augmentation Impacts}

When prompted not to rely on their inbuilt knowledge, LLM performance degraded compared to answering questions without access to the project source. For related questions, providing models with access to file-level and structural graph-based RAG improved the average score to \textbf{\textit{6.11}}, compared to file-based RAG only (\textbf{\textit{5.66}}). The results of the ablation study, shown in Table~\ref{tab:results_ablation},\footnote{Only Claude was used for the ablation study.} indicate that the structural Graph Agent contributes to the majority of the performance gains.

\begin{table}[ht]
    \centering
    \caption{Results from different configurations for ablation study}
    \label{tab:results_ablation}
    \rowcolors{1}{purple1!10}{white} 

    \resizebox{0.46\textwidth}{!}{
        \begin{tabular}{|c|c|c|}
            \hline
            \rowcolor{purple1!60}
            \color{white} \textbf{Configuration} & \color{white} \textbf{Score (related)} & \color{white} \textbf{Score (unrelated)} \\
            \hline
            File Agent only & 5.66 & 5.46\\ 
            \hline
            Graph Agent only & 6.12 & 5.49\\
            \hline
            \shortstack{All agents \\ \small{(Without LLM inbuilt knowledge)}} & 6.11 & 5.60\\
            \hline
            \shortstack{All agents \\ \small{(With LLM inbuilt knowledge)}} & 6.95 & 5.43 \\
            \hline
        \end{tabular}
    }
\end{table}

However, the performance of the multi-agent system when instructed to ignore inbuilt knowledge remained below that achieved when models answered using their inbuilt knowledge alone (i.e., without source access). This pattern suggests that LLMs have memorized many of the question--answer pairs during pretraining, a phenomenon widely documented in prior work on memorization and data extraction in large language models~\cite{carlini2021extracting}. As a result, performance declined when models were explicitly instructed to derive answers solely from retrieved source code.

When the same multi-agent configuration was paired with an LLM allowed to use both retrieved context and internal knowledge, performance on related questions improved to an average score of \textbf{\textit{6.07}}, significantly outperforming unrelated questions (\textbf{\textit{5.43}}, $p$-value $< 0.016$). Table~\ref{tab:results} summarizes these multi-agent results across different LLMs, showing consistent gains for related questions but limited improvement—or degradation—for unrelated ones.

When retrieval augmentation was applied to questions both before and after the model training cutoff, performance improved in both cases (Before: \textbf{\textit{7.00}} from \textbf{\textit{6.59}}; After: \textbf{\textit{5.68}} from \textbf{\textit{5.44}}). This indicates that augmentation generally has a positive effect, although performance on unseen questions remains moderate.

\begin{table}[ht]
    \centering
    \caption{Results of Multi-agent system on QA task using different LLMs}
    \rowcolors{1}{purple1!10}{white} 
    \label{tab:results}
    \begin{tabular}{|c|c|c|c|c|c|c|}
        \hline
        \rowcolor{purple1!60}
        \color{white} 
        \multirow{2}{*}{} & \multicolumn{2}{c|}{\color{white} \textbf{Related Questions}} & \multicolumn{2}{c|}{\color{white} \textbf{Unrelated Questions}}\\
        \cline{2-5}
        \rowcolor{purple1!60}
        \color{white} 
        \textbf{LLM} & \color{white} \textbf{Claude} & \color{white} \textbf{GPT} & \color{white} \textbf{Claude} & \color{white} \textbf{GPT} \\
        \hline
            Claude & 6.45  & 7.45  & 6.18 & 6.22 \\ 
        \hline
        GPT-4o  & 4.78  & 5.61  & 3.85 & 4.49 \\
        \hline
    \end{tabular}
\end{table}

Finally, comparing individual augmentation strategies shows that structural graph-based RAG achieves a higher average score (\textbf{\textit{6.12}}) than file-based RAG (\textbf{\textit{5.66}}, $p$-value $< 0.0002$), highlighting the importance of structural information for repository-level question answering. In contrast, introducing augmentation for unrelated questions reduced performance, consistent with recent findings that irrelevant or distracting context can harm LLM performance~\cite{yoran2024making,amiraz2025distracting}.

\begin{tcolorbox}[colback=purple1!10, colframe=purple2!60, title=Key Finding (RQ2)]
Retrieval augmentation improves performance, but gains depend strongly on the retrieval strategy. File-level RAG provides limited benefit, while graph-based RAG yields more consistent improvements, though performance on unseen (post-cutoff) questions remains modest.
\end{tcolorbox}

\section{Discussion}\label{sec:discussion}

Our study provides one of the first systematic evaluations of LLMs on repository-level question answering. The findings reveal important insights into both the potential and current limitations of LLMs for project-scale program comprehension.
\subsection{Overall Performance Gap}
The overall accuracy achieved by both GPT-4o and Claude 3.5 remains far below what would be required for practical deployment in developer workflows. Performance degraded markedly when models were prevented from relying on memorized training data, underscoring that much of their current success stems from pretraining exposure rather than genuine reasoning over source code. Fine-tuning LLMs on individual repositories is not a sustainable solution, as projects evolve continuously. This highlights the need for more robust RAG approaches that can dynamically adapt to unseen projects and unseen changes to projects.

These findings are particularly concerning for projects where few or no public Stack Overflow Q\&A pairs are available. For example in private or niche projects, models may have little or no training exposure. If LLM-generated answers are to reflect the current state of a codebase, retrieval mechanisms must be improved to capture up-to-date, contextually relevant information.

Our evaluation of the tool-only configuration, where LLMs relied exclusively on repository artifacts, further exposed the fragility of their reasoning. Without access to memorized Q\&A pairs, performance collapsed. These findings should caution practitioners against the use of LLMs for repository-level program comprehension. While humans often exploit structural cues such as dependency graphs, call hierarchies, or design patterns, LLMs fail to do so effectively without tailored retrieval mechanisms. This highlights that apparent success on public benchmarks may overstate true comprehension unless memorization effects are carefully controlled.

\subsection{Graph RAG as a Key Enabler}
Among the augmentation strategies, graph-based RAG consistently provided stronger improvements than file-level retrieval. By modeling structural relationships such as inheritance, containment, and cross-file calls, Graph RAG supplies the type of context developers rely on in practice. This aligns with Liu \et~\cite{liu2024codexgraph}, who reported that graph representations of structural data improved performance on code generation tasks.

Chen \et~\cite{chen_coreqa_2025} found that even with unlimited context length, LLMs fail to capture cross-file relationships when provided the entire codebase, resulting in poor QA performance. Our evaluation further confirms this limitation and underscores the need for systematic representation and retrieval techniques, such as structural dependency graphs, for effective repository-scale reasoning. We plan to investigate how additional structural information such as call graphs and data flow could impact performance.

In line with prior work, Claude 3.5 outperformed GPT-4o in leveraging graph-based context, suggesting that model-specific differences in handling structured outputs can affect multi-agent system effectiveness. This aligns with findings from~\cite{yanlin2024cypherbench}, which showed that Claude demonstrates higher proficiency in generating effective graph database queries. Future work could explore representing graph data in relational databases and compare this approach to the Neo4j graph database used in this study, in order to assess how well LLMs can generate correct Cypher queries versus SQL queries and its impact on question answering performance.

\subsection{Memorization vs. Reasoning}
Our cutoff-based analysis provides clear evidence that LLMs often succeed on repository-level QA tasks primarily due to memorization of Stack Overflow content. When models were prompted to ignore internal knowledge, accuracy dropped substantially, highlighting a methodological concern: evaluations that do not control for pretraining exposure risk overestimating LLM capabilities for program comprehension.

We further observed that some LLM responses were near verbatim reproductions of accepted Stack Overflow answers. We outline examples from our study with question IDs (QIDs) from our study. For example, in Debezium (QID:6887), the model reproduced the exact JSONB field access expression; in JUnit4 (QID: 8633), it closely mirrored the ground-truth explanation for covering the \texttt{next()} method; and in Ebean (QID: 10582), it repeated the instruction to replace a primitive \texttt{int} with the wrapper type \texttt{Integer}. Similarly, responses for OpenRouteService (QID: 9255) and WireMock (QID: 37860) echoed the accepted solutions almost verbatim, including distinctive technical phrasing such as ``Stateful Behaviour.'' These observations suggest that the models have likely encountered these question–answer pairs during training, aligning with recent concerns about benchmark contamination in LLM evaluation~\cite{hernandez2023scaling}. To our knowledge, this is the first empirical study to document such effects in the context of \textit{repository-level question answering}.

This evidence underscores the need for future benchmarks and evaluation protocols that explicitly disentangle memorization from reasoning. Promising approaches include cutoff-based dataset splits, unseen repository evaluation, augmentation techniques that encourage reasoning over code artifacts, and metrics assessing grounding in repository context. Based on our findings, \textit{we advocate for cutoff-based evaluation and analysis as a standard for future LLM studies in software engineering}.

\subsection{Retrieval Noise and the Distracting Effect}
While RAG holds promise, our experiments confirm that augmentation alone is insufficient for repository-level question answering, where relevant evidence is often buried among a large volume of unrelated or weakly related artifacts. Performance degraded notably when queries were supplemented with irrelevant context. In large repositories, multiple files, packages, and dependencies may be tangentially related to a query, sharply increasing the risk of retrieving distracting information. 

Prior work has shown that irrelevant or misleading passages can reduce model accuracy even when relevant content is present~\cite{yoran2024making,amiraz2025distracting}. This ``distracting effect'' highlights LLMs' vulnerability to noise, particularly from hard negatives, which can overwhelm reasoning and mislead the model. Techniques such as the Context Awareness Gate~\cite{heydari2024context}, which dynamically decide whether to incorporate retrieved context, illustrate one promising direction. More broadly, adaptive filtering and gating strategies will be essential to make retrieval augmentation robust in real-world software engineering scenarios, where contextual clutter is unavoidable.

\subsection{Impact of Question Type}

Analysis of scores across projects revealed that question type influences LLM performance. Though we did not perform a formal coding or analysis, we observed high-scoring responses were typically associated with \textit{API usage} or \textit{conceptual clarification}, while low scores were common for \textit{configuration}, \textit{integration}, or \textit{query-related} questions. For example, questions involving Lucene API (QID: 2986) usage were answered accurately, whereas tasks requiring Solr configuration (QID: 177649) or Hibernate integration often resulted in incomplete or misleading responses. These patterns indicate that LLMs handle stable, well-documented API tasks reasonably well but struggle with repository-level reasoning that depends on context-specific details.

This aligns with prior findings on developer comprehension: conceptual or API-focused tasks demand different strategies than procedural, configuration-heavy tasks~\cite{roehm_how_2012,latoza_importance_2010}, and configuration/environment questions are particularly prone to obsolescence~\cite{zhang2019empirical}. Similarly, LLMs are vulnerable when contextual clutter is high or reasoning requires synthesizing multiple, evolving artifacts~\cite{heydari2024context}. The contrast between API-level and configuration-heavy tasks suggests that repository-level QA is heterogeneous. Future experiments should consider question diversity, and practitioners should combine LLMs with retrieval and filtering strategies to mitigate risks in rapidly evolving or integration-heavy environments.



\section{Limitations and Future Work}
\subsection{Limitations}\label{sec:limitations}
This work has several acknowledged limitations. First, StackRepoQA is limited to Java projects, chosen due to their prevalence in large-scale systems; results may differ for other programming languages or ecosystems. Second, although Stack Overflow provides a rich source of real developer questions, it may not fully capture the breadth of challenges developers face in practice, particularly in industrial settings or private repositories. Third, our evaluation focuses on a limited set of augmentation strategies—specifically file-level and graph-based RAG—leaving open the possibility that alternative approaches, such as  specialized fine-tuning, could yield different results. 

\subsection{Threats to Validity}\label{sec:threats-to-validit}
Our study is subject to several threats to validity.
\textbf{Construct validity} may be affected by our reliance on accepted Stack Overflow answers as a proxy for correctness. While acceptance reflects community endorsement rather than definitive ground truth, this practice is consistent with prior empirical studies that leverage Stack Overflow data for large-scale evaluation. In addition, our use of LLM-as-a-judge evaluation may introduce bias, and while prompting models to ignore internal knowledge led to a clear performance drop, perfect compliance cannot be guaranteed.

\textbf{Internal validity} concerns arise from the dataset construction process. Keyword-based matching between Stack Overflow questions and GitHub repositories may still produce false positives despite manual filtering. Moreover, the choice of seven questions per project, while informed by power analysis, may not fully capture the diversity of developer questions associated with each repository.


\textbf{External validity} is limited by the scope of the dataset. StackRepoQA focuses on Java projects and Stack Overflow questions, which may not generalize to other programming languages, ecosystems, or industrial and private development settings. Our evaluation only incorporates two LLMs (Claude 3.5 Sonnet and GPT-4o), which may limit the generalizability of our findings across other model architectures, providers, or future generations of models.

\subsection{Future Work}\label{sec:future-work}
Our study suggests several directions for future work. First, many program comprehension questions lack canonical ground-truth answers~\cite{latoza_importance_2010,roehm_how_2012}. Future studies could examine how LLMs handle such open-ended questions through case studies or user-centered evaluations in realistic workflows, such as code reviews where repository context and history are essential~\cite{lin2025codereviewqa}.

Second, while our graph-based retrieval captures structural relationships, richer behavioral signals--such as data-flow and control-flow information, dependency resolution, or execution traces—may further support reasoning about program behavior and improve interpretability. Extending StackRepoQA beyond its current focus on Java to additional languages (e.g., Python, C\#, JavaScript), as well as incorporating other collaboration artifacts such as GitHub Issues, Discussions, or additional Stack Exchange communities, would enable evaluation across more diverse ecosystems and settings.

Finally, integrating non-textual representations, including visual or IDE-style views such as class diagrams, call graphs, or navigation aids, may further enhance repository-scale comprehension by combining structured views with LLM-driven question answering.


\section{Conclusion}\label{sec:conclusion}
This paper introduced StackRepoQA, the first publicly available, multi-project, cutoff-aware benchmark for repository-level question answering, comprising $1{,}318$ real Stack Overflow questions mapped to $134$ actively maintained Java projects. Using this dataset, we evaluated state-of-the-art LLMs on project-scale comprehension tasks and observed three key findings: (1) current models achieve only moderate accuracy, with much of their apparent success attributable to memorization rather than reasoning over source code; (2) retrieval augmentation improves performance, but naive file-level retrieval is insufficient, while graph-based retrieval that captures structural relationships yields more robust gains; and (3) performance degrades sharply on questions posted after model training cutoffs, underscoring the importance of benchmarks that disentangle memorization from reasoning.

Together, these results highlight both the potential and the limitations of current LLMs for repository-scale program comprehension. StackRepoQA provides a reproducible foundation for studying LLM–code interactions beyond snippet-level tasks, while our findings caution against relying on LLMs as stand-alone solutions for private or rapidly evolving repositories without structured retrieval and verification mechanisms. By releasing the dataset and evaluation pipeline, we aim to support future research on trustworthy, structure-aware LLM systems for real-world software development.


\bibliographystyle{acm}
\bibliography{reference}
\end{document}